\documentclass[showpacs,preprintnumbers,amsmath,amssymb,twocolumn,floatfix,nofootinbib]{revtex4-1}
\usepackage[usenames]{color}
\usepackage{epsfig,multirow}
\usepackage{amssymb}
\usepackage{xcolor}
\newcommand{\be}{\begin{equation}}
\newcommand{\ee}{\end{equation}}
\newcommand{\bee}{\begin{eqnarray}}
\newcommand{\eee}{\end{eqnarray}}

\definecolor{grey}{rgb}{0.9,0.9,0.9}
\definecolor{black}{rgb}{0,0,0}

\def \irbaddress{Rudjer Bo\v{s}kovi\'{c} Institute, Bijeni\v{c}ka cesta 54, P.O. Box 180, 10002 Zagreb, Croatia}
\def \untzaddress{University of Tuzla, Faculty of Science, Univerzitetska 4, 75000 Tuzla, Bosnia and Herzegovina}
\def \mainzaddress{Institut f\"{u}r Kernphyik, Universit\"{a}t Mainz, D-55099 Mainz, Germany}
\def \GWUSAIDaddress{Data Analysis Center at the Institute for Nuclear Studies, Department
of Physics, The George Washington University, \vspace*{0.5cm} Washington, D.C. 20052 }

\begin{document}

\title{Introducing Pietarinen expansion method into \\ single-channel pole extraction problem}
\author{Alfred \v{S}varc}
\email{alfred.svarc@irb.hr}
\affiliation{\irbaddress}
\author{Mirza Had\v{z}imehmedovi\'{c}}
\affiliation{\untzaddress}
\author{Hedim Osmanovi\'{c}}
\affiliation{\untzaddress}
\author{Jugoslav Stahov }
\affiliation{\untzaddress}
\author{ Lothar Tiator}
\affiliation{\mainzaddress}
\author{Ron L. Workman}
\affiliation{\GWUSAIDaddress}

\date{\today}

\begin{abstract}
  We present a new approach to quantifying pole parameters of single-channel processes based on a Laurent expansion of 
partial-wave T-matrices in the vicinity of the real axis. Instead of using the conventional power-series description of the 
non-singular part of the Laurent expansion, we represent this part by a convergent series of Pietarinen functions. 
As the analytic structure of the
non-singular part is usually very well known (physical cuts with branch points at inelastic thresholds, and unphysical cuts in 
the negative energy plane), we find that one Pietarinen series per cut represents the analytic structure fairly reliably. The 
number of terms in each Pietarinen series is determined by the quality of the fit.  The method is tested in two ways: on a toy 
model constructed from two known poles, various background terms, and two physical cuts, and on several sets of realistic 
$\pi$N elastic energy-dependent partial-wave amplitudes (GWU/SAID - \cite{GWU,GWU1}, and Dubna-Mainz-Taipei - \cite{DMT0,DMT}).  
We show that the method is robust and confident using up to three Pietarinen series, and is particularly 
convenient in fits to amplitudes, such as single-energy solutions, coming more directly from experiment; 
cases  where the analytic structure of the regular part is a-priori unknown. 

\end{abstract}

\maketitle

\section{Introduction}

Revisions to the Review of Particle Properties \cite{PDG} and contributions to the recent Camogli workshop \cite{Camogli2012}, 
among others, have emphasised the fact that poles, and not Breit-Wigner parameters, determine and quantify resonance 
properties, and that they should be used as a link between scattering theory and QCD. However, at the same time, the question 
of finding an adequate procedure to extract pole parameters 
from single-channel T-matrices remains open. Experimentalists are quite familiar with fits to data using Breit-Wigner functions 
(either with constant parameters and very general backgrounds, or with energy dependent masses and widths), but are less 
experienced when complex energy poles are to be used. Simple procedures for pole extraction exist, but are not reliable in all 
cases. 
At present, poles are usually extracted from theoretical single or multi-channel models, fitted to the data, using an array of 
standard pole extraction methods: analytic continuation of the model functions into the complex energy plane 
\cite{Doering,EBAC,CMB,Zagreb,Bonn}, speed plot \cite{Hoehler93}, time delay \cite{Kelkar}, N/D method \cite{ChewMandelstam}, 
regularization procedure \cite{Ceci2008}, etc. However, this typically requires 
solving an involved
single/coupled-channel model and analyzing the obtained analytic solution, which implicitly contains
both parts: singular and background. Hence, the analytic form of the full solution varies from model to model; the 
pole-background separation method is not unique, and requires an intimate knowledge of the underlying model.  The intention of 
this paper is to offer a simple, robust and confident method to obtain scattering-matrix poles, while avoiding the use of any 
particular
assumptions concerning the form of background terms. 
We base our analysis on a Laurent expansion of partial wave T-matrices, which uniquely separates singular from finite terms, 
and treat singular and finite terms separately. Our main assumption is that all scattering matrix poles are of the first order.

\section{Formalism}

\subsection{Laurent expansion}

We start with a fact that each single-channel scattering matrix is a fully analytic function in the complex energy plane with a 
well defined number of poles and cuts, and all knowledge from complex analysis may be applied directly to it.  
Consider first the Laurent expansion of a complex analytic function, $f(z)$. At the most basic level, we have the statement:

\noindent
 \emph{Suppose that f is holomorphic in the annulus \{$ z \in  \mathbb{C}$, $ R_1 < |z - z_0| < R_2$\}, where $0 \leq R_1 < R_2 
<\infty$. Then we can write f as a Laurent series: for $R_1 < |z - z_0| < R_2$ we have}
\begin{eqnarray}
f(z) &=&  \sum _{n=0}^{\infty} a_n (z-z _0)^n + \sum _{n=1}^{\infty} a_{-n} (z-z _0)^{-n}
\end{eqnarray}
This theorem basically says that each complex function defined in the full complex energy plane can \emph{locally} be 
represented by a power series. Observe that this is not an identity in the full complex energy plane, the \emph{globally} 
unknown, complicated function $f(z)$ can \emph{locally} (in the vicinity of the singularity) be represented by a simpler 
function consisting of energy independent pole contribution(s) and a usually limited number of power series terms. 
 
Applied to a single channel scattering matrix in the vicinity of one simple pole at $\omega = \omega_ 0$  in the complex energy 
plane $\omega$, adjusted to the sign conventions of PDG \cite{PDG}, we have: 
\begin{eqnarray}
\label{eq:Laurent}
T(\omega) &=& \frac{a_{-1}}{ \omega_0-\omega}+ \sum _{n=0}^{\infty} a_n (\omega _0-\omega)^n;    \nonumber \\ & & \, \, \, \, 
a_{n}, \, \omega, \, \omega_0 \in  \mathbb{C}. 
\end{eqnarray}

This expression is simply the generalization of a Taylor's expansion of an analytic function about any regular point to the 
expansion of singular function around a singular point, $\omega_0$, 
and as such is a unique representation of the function over a part of the complex energy plane defined by its radius of 
convergence. 

However, the functions we consider may (and do) contain more than one pole for $\omega \neq \omega_ 0$. If we iterate this 
procedure using the Mittag-Leffler theorem \cite{Mittag-Leffler}, which says that a meromorphic function can be expressed in 
terms of its poles and associated residues combined with an additional entire function,  we can, without loss of generality, 
write down the generalized Laurent expansion for the function with $k$ poles:
\begin{eqnarray}
\label{eq:Mittag-Leffler}
T(\omega) &=& \sum _{i=1}^{k} \frac{a_{-1}^{(i)}}{\omega_i-\omega }+B^{L}(\omega);  \nonumber \\  
& & \, \, \, \, a_{-1}^{(i)}, \omega _i, \omega \in  \mathbb{C},
\end{eqnarray}
where $k$ is number of poles. $a_{-1}^{(i)}$ and $\omega_ i$ are residua and pole positions for i-th pole respectively, and 
$B^{L}(\omega )$ is a function regular in all $\omega  \neq \omega _i$.
\noindent
The most general starting point, for the functions which we consider, is then Eg.~(\ref{eq:Mittag-Leffler}).
\begin{center}
\emph{Is this a Breit-Wigner expansion with constant coefficients?} 
\end{center}

At a first glance, one might be tempted to miss-identify Eq.~(\ref{eq:Mittag-Leffler}) as the oldest, simplest, and generally 
inapplicable, expansion of a single-channel scattering matrix in terms of Breit-Wigner functions with constant coefficients:
\begin{eqnarray}
\label{eq:Breit-Wigner}
T(\omega) & \approx & \sum _{i=1}^{k} \frac{x_i \, \Gamma_i/2}{ M_i-\omega - \imath \Gamma_i/2}+B(\omega) \nonumber \\
& & \, \, \, \, x_i, \ M_i, \, \Gamma_i, \omega   \in  \mathbb{R},
\end{eqnarray}
where $M_{i}$ is resonance mass, and $x_{i}$ and $\Gamma_{i}$ its elasticity and width.
We emphasize that this is incorrect.

Clearly, the analytic structure of right-hand sides of Eqs.~(\ref{eq:Mittag-Leffler}) and (\ref{eq:Breit-Wigner}) are 
different. Eq.~(\ref{eq:Mittag-Leffler}) has a complex coefficient associated with the singular term, while in 
Eq.~(\ref{eq:Breit-Wigner}) it is real. The presence of a complex coefficient for the singular part of 
Eq.~(\ref{eq:Mittag-Leffler}) fundamentally changes the form of the real and imaginary parts of the investigated functions, so 
these two functions are not mutually interchangeable.  
Beyond this, the essential fact which distinguishs the Breit-Wigner expansion with constant coefficients - Eq. 
(\ref{eq:Breit-Wigner}) from a Laurent expansion - Eq.(\ref{eq:Mittag-Leffler}) \emph{is the domain of applicability}.

When the Breit-Wigner expansion with constant coefficients is traditionally employed, it is assumed that the function is 
defined everywhere in the complex energy plane. This, however, raises important problems, such as the existence of poles on the 
first, physical sheet, which is strictly forbidden.
The Laurent expansion, however, is valid only locally. 
\\ \\ \noindent
Let us illustrate this with an example.
\\ \\ \noindent
To emphasize the point, we note that even the most general energy-dependent Breit-Wigner function can be locally represented 
by its Laurent expansion:
\begin{eqnarray}
\label{eq:Breit-Wigner ED}
\frac{\Gamma(\omega)e^{\imath \phi}}{M(\omega)- \omega- \imath K(\omega)\Gamma(\omega)} & 
\equiv&\frac{a_{-1}}{\omega_0-\omega}+ \sum _{n=0}^{\infty} a_n (\omega _0-\omega)^n. \nonumber \\
\end{eqnarray}
However, the left-hand side is a function valid in the whole complex energy plane, with poles (when the function is properly 
defined) only on the unphysical sheet and higher Riemann sheets, while the right-hand side is a function defined only locally, 
in the vicinity of poles. Outside the radius of convergence this function diverges. 

\begin{figure}[!t]
\includegraphics[width=0.9\columnwidth]{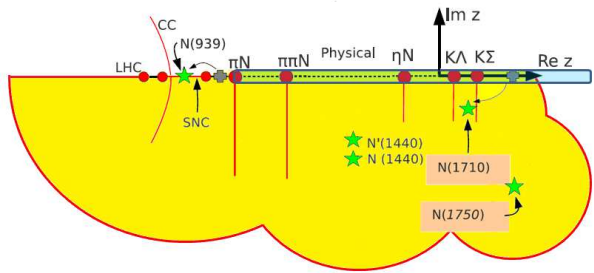}  
\caption{\small(Color online) Domain of convergence of Laurent series for P$_{11}$ partial wave.}
\label{JuelichP11} 
\end{figure}

As an illustration of this feature, we give the area of convergence for the P$_{11}$ partial wave of J\"{ulich} model  
\cite{Ronchen2012} as seen by Laurent method in Fig.~(\ref{JuelichP11}).  (Picture is produced on the basis of Fig.~(39) of 
this reference.)  The domain of convergence for the Laurent expansion of the J\"{ulich}  P$_{11}$ amplitude is indicated in 
yellow, while the J\"{ulich}  solution (possibly unknown to us), is valid over the full complex energy plane. We see that the 
effective area of convergence of the Laurent method is much smaller than the area of convergence of the original model. It is, 
however, sufficient to enable the analytic continuation of data from the real axis to the nearest singularities. Our aim is not 
to
reproduce the full function, but only the pole parameters of nearby singularities.

 \subsection{Pietarinen series} 

In the following, we propose the use of an expansion, differing from the standard power series
for the regular part of Laurent expansion given in Eq.(2). The preferred conformal mapping expansion variable, giving what we 
call the
Pietarinen series, has been proposed and introduced by
Ciulli \cite{Ciulli,Ciulli1} and Pietarinen \cite{Pietarinen}, and it
has been used, with great success, in the Karlsruhe-Helsinki partial wave analysis \cite{Hoehler84} where 
invariant amplitudes have been expanded in as many as 50 terms.  The aim of this article is not  to give the solution in the 
full complex energy plane which is matching the data, but to give only a locally valid representation of the exact solution, 
which is however enabling us to extract poles using the data from the real axes.
 The novelty of our approach is that we propose to avoid discussing the arbitrariness of all possible choices for the 
background function $B^{L}(\omega )$ by replacing it with rapidly converging Pietarinen  power series defined by a complete set 
of functions with well known analytic properties. 

If $F(\omega)$ is a general, unknown analytic function having a cut starting at $\omega=x_P$, then using a conformal mapping 
technique, it can be represented as a power series, having the maximum radius of convergence, in the following way:
\begin{eqnarray}
\label{eq:Pietarinen}
F(\omega ) &=& \sum_{n=0}^{N}c_n\, Z^n(\omega ), \, \, \, \, \, \, \, \, \, \, \omega  \in  \mathbb{C}   \nonumber \\
Z(\omega )&=& \frac{\alpha-\sqrt{x_P-\omega }}{\alpha+\sqrt{x_P-\omega }}, \, \, \, \, \, c_n, x_P, \alpha \in  \mathbb{R},
\end{eqnarray} 
with the $\alpha$ and $c_n$ being tuning parameter and coefficients of Pietarinen function $Z(\omega)$ respectively.  

The essence of the approach is the fact that a set $(Z^n(\omega ), \, n=1, \, \infty)$ forms a complete set of functions 
defined on the unit circle in the complex energy plane having branch cut starting at $\omega= x_P$, but the analytic form of 
the function is not otherwise defined. The final form of the analytic function $F(\omega)$ is obtained by introducing the 
rapidly convergent power series with real coefficients, and the degree of the expansion is determined in fitting the input 
data. 

\subsection{The application of Pietarinen series to scattering theory}

Each partial wave contains poles which identify resonant contributions, and has cuts in the physical region starting at 
thresholds of elastic and all possible inelastic channels, and finally there are
$ t $-channel, $u$-channel and nucleon exchange contributions quantified with corresponding negative energy cuts. 
However, the explicit analytic form of each cut contribution is not known. Instead of guessing the exact analytic form of all 
of these, we propose to use one Pietarinen series to represent each cut. The number of terms in each Pietarinen series will be 
determined by the quality of fit to the input amplitudes. 
In practice, we have too many cuts (especially in the negative-energy range), so we reduce their number by dividing them in two 
categories: all negative energy cuts are approximated with only one, effective negative energy cut represented with one 
Pietarinen series, and each physical cut is represented with its own Pietarinen series with branch points determined by the 
physics of the process. 
  
In summary, the set of equations which define Laurent expansion + Pietarinen series method (L+P method) is:
\begin{eqnarray}
\label{eq:Laurent-Pietarinen}
T(\omega ) &=& \sum _{i=1}^{k} \frac{a_{-1}^{(i)}}{\omega_i-\omega}+ B^{L}(\omega)  \\
B^{L}(\omega)&=& \sum _{n=0}^{M}c_n\, Z^n(\omega )  +  \sum _{n=0}^{N}d_n\, W^n(\omega ) + \cdot \cdot \cdot    \nonumber  \\
Z(\omega )&=& \frac{\alpha-\sqrt{x_P-\omega}}{\alpha+\sqrt{x_P - \omega }}; \, \, \, \, \,   W(\omega ) =  
\frac{\beta-\sqrt{x_Q-\omega }}{\beta+\sqrt{x_Q-\omega }} + \cdot \cdot \cdot \, \, \,  , \nonumber
\end{eqnarray}
\noindent
where  $a_{-1}^{(i)}, \omega _i, \omega  \in   \mathbb{C}$,  $c_n, x_P, d_n, x_Q, \alpha, \beta \in  \mathbb{R}$ and ${\rm and} 
\, \, \, k, M, N \in  \mathbb{N}$.
All parameters in set the of equations (\ref{eq:Laurent-Pietarinen}) are determined by the fit. As our input data are on the 
real axis, the fit is performed only on this dense subset of the complex energy plane. 

\subsection{Fitting procedure}

The class of input functions which can conveniently to be analyzed with this method is quite wide. 
One may either fit partial-wave amplitudes obtained from any theoretical model, or experimental data directly. In either case, 
the T-matrix is represented by the given set of equations (\ref{eq:Laurent-Pietarinen}), and a minimization 
criterion (the discrepancy parameter) with respect to the input data is defined (usually $\chi^2$ type), and a fit is done.
As the partial wave $T$ given by equations (\ref{eq:Laurent-Pietarinen}) does not fulfill the unitarity condition manifestly, 
we  impose elastic unitarity below the first inelastic threshold numerically by introducing a penalty function into the  
discrepancy parameter of the fit.

We start with minimal number of poles, three Pietarinen functions (one for background, and two for dominant physical channels),  
and a minimal number of Pietarinen fitting parameters $c_n, \, d_n \, ...$ . We usually start with $M, \,N=5$.

\noindent
The discrepancy parameter is defined as:
\begin{eqnarray}
\label{discrepancy}
\chi^2 & =& \sum _{j=1}^{N_{pts}} \mid T^{inp}(\omega _j) - T(\omega _j) \mid ^2  /{\rm w}_{j}^2 + \sum_{j=1}^{3}\lambda^j \, 
\chi_{Pen}^j + \nonumber \\
        & + &  \,  \Upsilon \, \, \sum_{j=1}^{N_{pts}^{el}}(1-S(\omega_j)S(\omega_j)^\dagger)^2 ,  
\end{eqnarray}
  where  ${\rm w}_{j}$ is the corresponding statistical weight (standard error bar for experimental data and adequate 
uncertainty parameter for theoretical functions) and $\chi_{Pen}^j =  \sum _{k=1}^{N} (c_k^j)^2 \, k^3$ is the Pietarinen 
penalty function \cite{Pietarinen} which guarantees the soft cut-off of higher order terms in the Pietarinen expansion. The 
third term in Eq. (\ref{discrepancy}) is introduced to impose elastic unitarity as discussed above.  Parameters $\lambda^j$ and 
$\Upsilon$ are penalty function adjusting parameters which serve to bring into correct proportion contributions from penalty 
functions with the contribution originating from the function itself. They are determined empirically, prior the fit, 
independently for each penalty function.

 The discrepancy parameter is analyzed, and the quality of the fit is also visually inspected by comparing the fitting function 
with fitted data. If the fit is unsatisfactory (the discrepancy parameter is high, or the fit visually does not reproduce the 
fitted data), the number of Pietarinen parameters $c_n, \, d_n \, ...$ is increased by one. The fit is repeated, and the 
quality of the fit is re-estimated. This procedure is continued until we have reached a sufficient number of Pietarinen terms 
so that we are able to reproduce the input data. If the quality of the fit is still unsatisfactory, we first increase the 
number of poles and then repeat the procedure.
 
\subsection{Limitations of the method}

By construction it is clear that the method has its natural limitations. As seen from set of Eqs. (\ref{eq:Laurent-Pietarinen}) 
our Laurent decomposition contains only two branch points in the physical region, and as seen from Fig.\ref{JuelichP11} this is 
far from enough in a realistic case. Any realistic analytic function in principle containing more than two branch points will 
in our model be approximated by a different analytic function containing only two. So, this will be the main source of our 
errors.

Error analysis will be discussed in detail, in forthcoming publications, when we extract pole positions from experimental data.

\section{Testing the method} 

 Being aware of the limitations of our method (limited number of branch points), we tested the method in two ways.
 
First we have tested the method on an elaborate toy model; a model in which we have full control over the analytic structure of 
the input. We have constructed a toy-model function imitating physical reality as close as possible (known pole parameters, two 
positive energy cuts and a background contribution), generating the input data table, and verified how well our method 
reproduces the known input parameters.

However, to prove the applicability of the method in reality, we have tested it on realistic amplitudes as well. We have used 
our L+P method to extract pole parameters first from several GWU/SAID energy dependent (GWU/SAID ED) $\pi$N elastic partial 
wave amplitudes with known (published) pole positions \cite{GWU, GWU1}, and also on the Dubna-Mainz-Taipei energy dependent 
(DMT ED)  \cite{DMT0,DMT} amplitudes. Exploring GWU/SAID pole structure is the first analysis of these amplitudes by the 
present collaboration, while Zagreb-Mainz-Taipei collaboration have already used DMT amplitudes for testing other possible 
single channel, but local, pole extraction methods such as the speed plot (SP), renormalization method, etc. For further 
details 
we refer the reader to reference \cite{DMT0,DMT}. We consider it important to mention that DMT amplitudes have more poles than 
have been reported in \cite{DMT0,DMT}, and we use them in our analysis as well \cite{DMT1}. Included here are the S$_{11}$, 
S$_{31}$  P$_{11}$, D$_{13}$, D$_{33}$, F$_{15}$ and F$_{37}$ amplitudes.  We expect that the agreement with "exact" results, 
obtained by analytic continuation of known functions into the complex energy plane, will be poorer than in the toy-model case, 
as each model has its own complicated analytic structure, and fitting with two branch points only will definitely influence 
pole positions. However, we show that the results in these cases are very good as well.

Finally let us also mention that analytical continuations of K-matrix and/or dynamical models are also not absolutely free of 
uncertainties. There are various elaborate numerical methods, which are not always free of numerical instabilities.
This is especially true when poles are very far away from the physical axis or when the structure of Riemann sheets in the 
models becomes complicated.

\subsection{Fitting the toy-model}

\begin{table*}[t!]
 \caption{Toy model parameters and fitted parameters. Input parameters are given
in boldface, and results of a fit in normal font. Table is given in GeV units. $\alpha, \beta$ , and $\gamma$ are tuning 
parameters in corresponding Pietarinen functions. \\ }
\label{tb1:Toy model parameters} 
\begin{tabular}{||c|cccccccc|ccc|ccc|ccc|c||}
\hline
   & $r_{1}$  & $g_{1}$  & $M_{1}$  & $\Gamma _1$  & $r_{2}$  & $g_{2}$  & $M_{2}$  & $\Gamma _2$  & $\alpha$  & $x_{P}$ & 
$N_{1}$ & $\beta$& $ x_{Q}$&$ N_{2}$  & $\gamma $ & $ x_{R}$ &$ N_{3}$  & $10^{2}\chi_{R}^{2}$\tabularnewline
\hline 
\hline 
\multicolumn{1}{||c|}{Toy-model} & \multicolumn{18}{c||}{}    \tabularnewline
\hline 
\multicolumn{1}{||c|}{} & \textbf{ 0.1}  & \textbf{0.09 }  & \textbf{1.65 }  & \textbf{0.165}  & \textbf{0.09}  & \textbf{0.06 
}  & \textbf{2.25}  & \textbf{0.2}  & \multicolumn{3}{c|}{} & \multicolumn{3}{c|}{} & \multicolumn{3}{c|}{} & 
\multicolumn{1}{|c||}{} \tabularnewline
\hline 
\hline 
\multicolumn{1}{||c|}{Fitted results} & \multicolumn{18}{c||}{}    \tabularnewline
\hline
  P1  & 0.085 & 0.102 & 1.663 & 0.171 & 0.087 & 0.075 & 2.262 & 0.216 & 1.09 & -2.64 & 10 &  & && &&& 328.19\tabularnewline
\hline
  P2  & 0.098 & 0.086 & 1.650 & 0.161 & 0.095 & 0.058 & 2.247 & 0.199 & 0.44 & -0.47 & 9 & 1.95 &3.97& 8 & && & 
70.37\tabularnewline
\hline
 P3  & 0.099 & 0.090 & 1.650 & 0.164 & 0.089 & 0.061 & 2.251 & 0.200 & 4.19 & -22.99 & 5 & 2.22& 3.98&5 & 1.67& 0.97& 3 & 
0.24\tabularnewline
\hline \hline
\end{tabular}
\end{table*}

In principle, we could have defined toy-model input data $T^{ty}(\omega _j)$ by defining a toy-model function and by  normally 
distributing its values in order  to simulate the statistical nature of real measured data. However, as the main goal of this 
paper is to establish the validity of the approach, we have restricted our analysis to infinitely precise data by using 
non-distributed toy-function values, and using a statistical weight ${\rm w}_j$ of 5 \%. 
\\ \indent
Our toy-model function is constructed by assuming a typical analytic structure of a partial wave: it is constructed as a sum of 
two poles, two physical cuts and several non-resonant background contributions.  The function representing physical cuts  is 
constructed from a function $f(x,a) = \sqrt{x^2-4 a x}/2x$ having a cut starting from $x_0= 4 a$  on the real axes\footnote{The 
type of the function used for physical cuts comes from the phase space factor for two body reactions 
$\phi(s)=\sqrt{\Lambda(s)}/2s$, with $\Lambda(s)=s^2 - 2 s (M^2+m_\pi^2)+(M^2-m_\pi^2)^2$, and taking $m_\pi=M$.}, and the 
analyticity is imposed through the once subtracted dispersion relation 
\mbox{$\Phi(x,a)=\frac{x-x_0}{\pi}\int_{x_0}^{\infty}\frac{f(x',a)}{(x'-x)(x'-x_0)} \, dx'$}.
However, to simplify our demonstration of utility of the L+P method, we replace all negative energy cuts with two poles deep in 
unphysical region. In spite of appearing rather restrictive, such an approximation is reasonably justified. Namely, we know 
that each cut can be represented by the infinite sum of poles, and as negative cut is indeed very far from the region of 
interest, replacing it with only two out of infinite number of poles is a good approximation (see Cutkosky CMB approach 
\cite{CMB}). 

Thus our toy-model function is given as: 
\begin{eqnarray}
T^{ty}(\omega )  &=& \sum _{k=1}^{ 2} \frac{r_{k} + i \, \, g_{k}}{ M_{k}-\omega  - i \, \, \Gamma_{k}/2} +  \\ 
                 &+ &  \Phi(\omega,0.25)+  \Phi(\omega,1.)+ B(\omega ),  \nonumber  \\
 \Phi(\omega,a)&=& \frac{\sqrt{\omega (-4 a + \omega)}}{2 \pi \omega} \ln  \frac{{2 a - \omega - \sqrt{\omega(-4 a + 
\omega)}}}{2a} \nonumber \\
B(\omega ) & = &  \frac{10.}{ -10. -\omega - i \, \,  5.}+   \frac{10.}{ -6. -\omega - i \, \, 4.}, \nonumber 
\end{eqnarray}
where $r_{k},g_{k},M_{k},\Gamma_{k} \in \mathbb{R}$. 

Toy-model parameters for all test cases are chosen to resemble physical reality as much as possible, and are given in Table 
\ref{tb1:Toy model parameters} with bold face characters. By construction,  our toy-model function has two poles and two cuts 
producing clearly visible pole structure and two pronounced cusps at 1.0 GeV ($a=0.25$ GeV) and 4.0 GeV ($a=1.00$ GeV). 

\begin{figure}[!h]
\includegraphics[width=0.9\columnwidth]{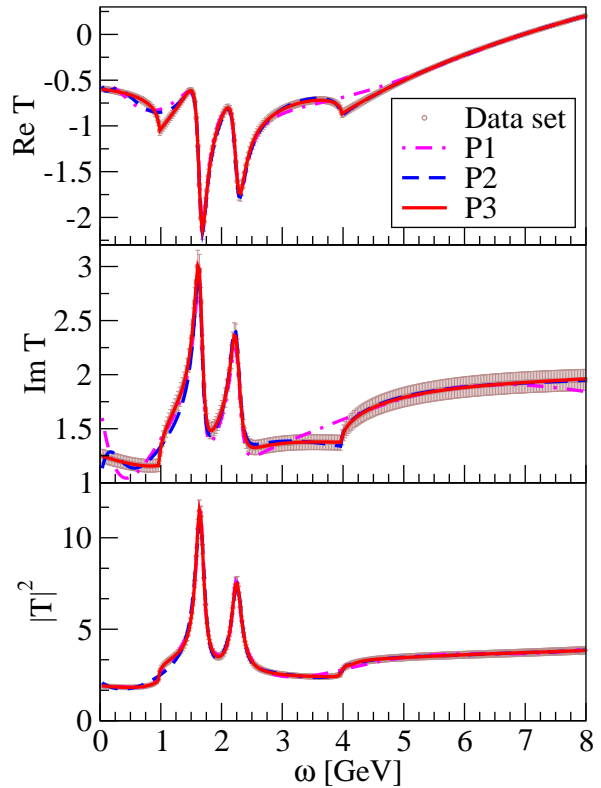}  
\caption{\small(Color online) Toy-model function. Dashed-dotted, dashed and full lines P1, P2 and P3  give the quality of the 
fit for solutions with one, two and three Pietarinen expansions respectively. 
 }
\label{Fig1} 
\end{figure}

Results of the fit for all cases are given in  Table~\ref{tb1:Toy model parameters} and Fig.~\ref{Fig1}.  For more numerical 
parameters we refer the reader to the reference \cite{Svarc2012} where an expanded version of the toy-model is given. 
\\ \\ \noindent
In conclusion, we claim that the basic assumptions of the approach have been confirmed. All pole parameters are almost exactly 
reproduced. The Pietarinen branch point $x_P$ turns out to be subthreshold  and effectively reproduces background 
contributions, and Pietarinen branch points $x_Q$ and $x_R$ correspond to the branchpoints of the toy-model function (1. and 
4.).

We have also tested the dependence of the fitting success upon the number of Pietarinen series used. 

In Table~\ref{tb1:Toy model parameters} and Fig.~\ref{Fig1}, we show a fit the full model only with one Pietarinen (P1), two 
Pietarinens (P2) and three Pietarinen series (P3). It is straightforward to see that if one uses only one expansion,  reduced  
$\chi^2$ ($\chi^2_{R}$ -  discrepancy parameter per degree of freedom)
is high (3.2819), and the fitting function "misses" both cusps. When the number of Pietarinen series raises to two (P2),  
reduced $\chi^2$ is improved (0.7037) and one cusp (higher one at 4) is covered, so we indeed need all three Pietarinen series 
(P3) to reproduce all poles and all cusps and obtain the best reduced $\chi^2= 0.0024$. We note that the very small $\chi^2$ is 
a measure of the goodness of our fit only. It is not a $\chi^2$ in a statistical sense, because the input "data" are not 
statistically distributed with standard error definition, but taken from a numerical function with an arbitrary error of 5\% 
being assumed.  In the final case, where the number of Pietarinen series corresponded to the number of cuts, all pole 
parameters perfectly corresponded to the toy-model input values.
\newpage
\subsection{Fitting the realistic input}

The L+P method allows the extraction of poles from partial waves obtained as result of theoretical analysis (''theoretical 
input'') and, we have used it to extract poles from GWU/SAID and DMT $\pi$N elastic amplitudes. 
However, if the ''theoretical input'' has analytic structure in the regular part different from the cut structure assumed in 
the fit, errors in fitting the background will be transferred to pole parameters, as the fit will try to compensate. \\ \\ 
\noindent
We apply the following strategy:\\ \\
1) We fix the  pion-nucleon elastic threshold position at the experimental value, and \\
2) Allow other threshold positions to vary as a parameters in a fit and find  optimal values which replace exact physical 
thresholds.

\begin{figure*}[!p]
\includegraphics[width=0.9\textwidth]{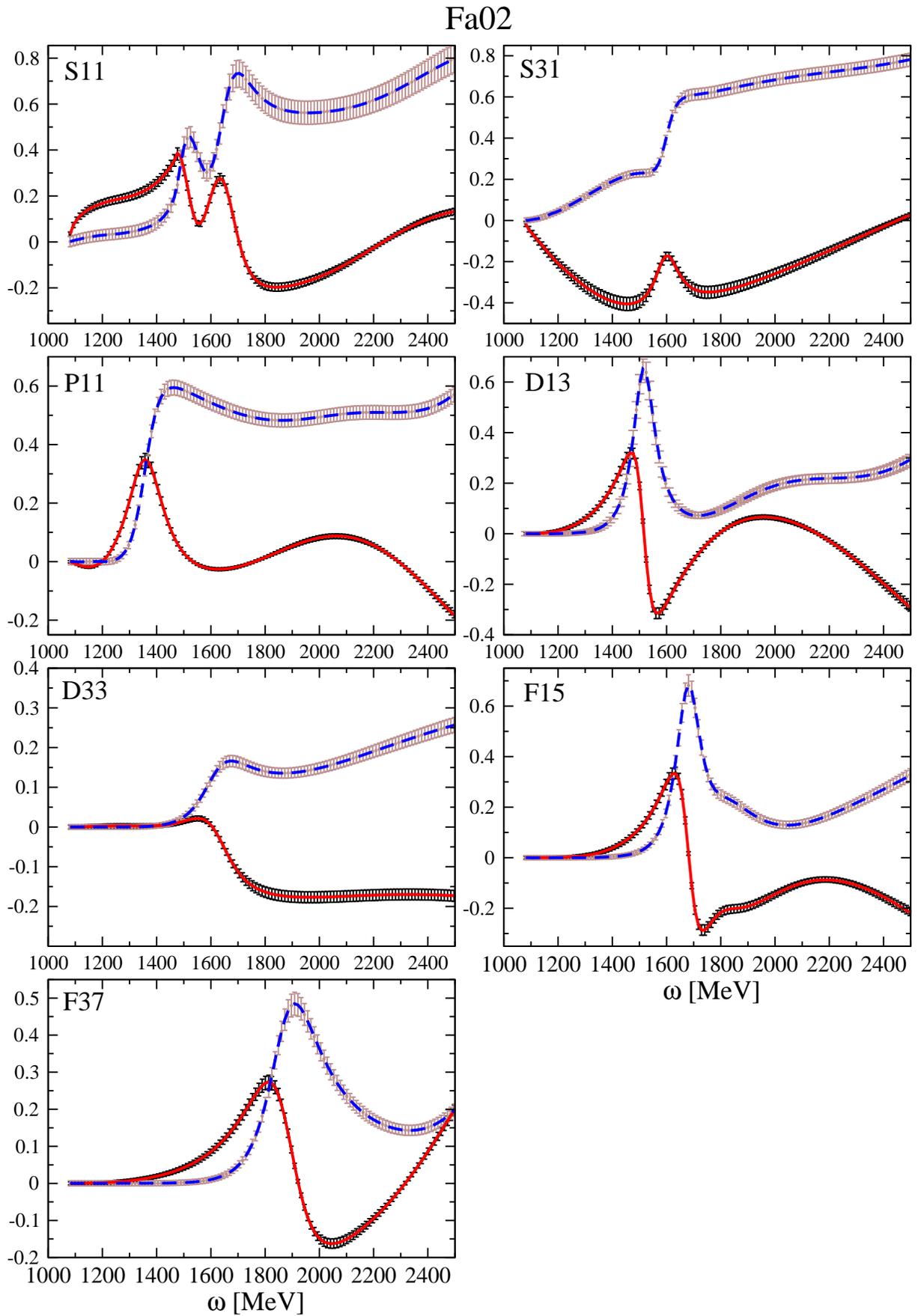}  
\caption{\small(Color online) L+P fit to partial waves from Fa02 solution. 
Dashed blue line and solid red line show fit to real and imaginary parts of partial waves, respectively.}
\label{Fa02_fig} 
\end{figure*} 
\begin{figure*}[!p]
\includegraphics[width=0.9\textwidth]{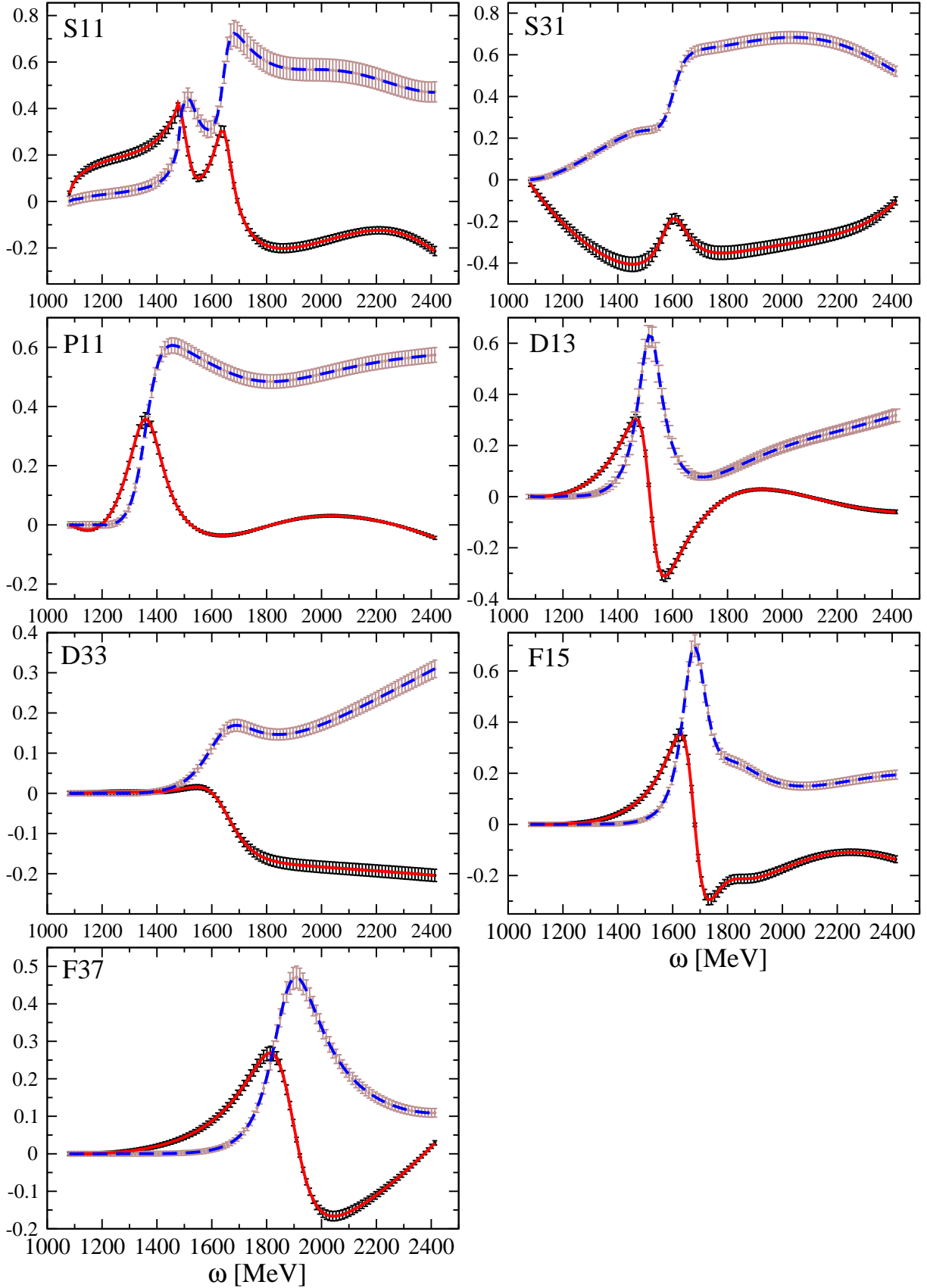}  
\caption{\small(Color online) L+P fit to partial waves from Sp06 solution.
 Dashed blue line and solid red line show fit to real and imaginary parts of partial waves, respectively.}
\label{Sp06_fig} 
\end{figure*} 
\begin{figure*}[!p]
\includegraphics[width=0.9\textwidth]{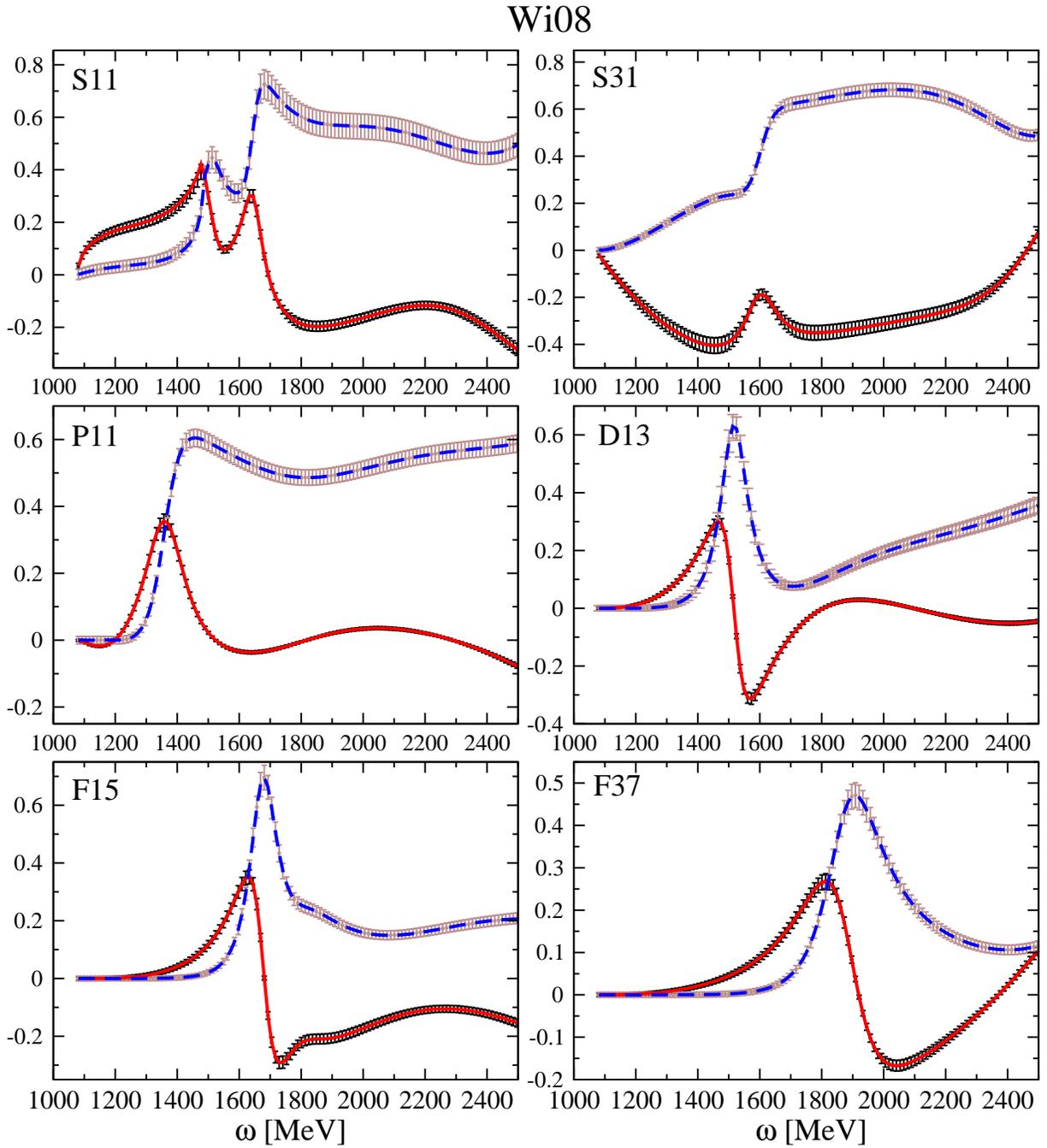}  
\caption{\small(Color online)  L+P to partial waves from Wi08 solution. For D$_{33}$ partial wave we have no values for poles, 
so we omitted it from fitting.
 Dashed blue line and solid red line show fit to real and imaginary parts of partial waves, respectively.}
\label{Wi08_fig} 
\end{figure*}

\begin{figure*}[!h]
\includegraphics[width=0.9\textwidth]{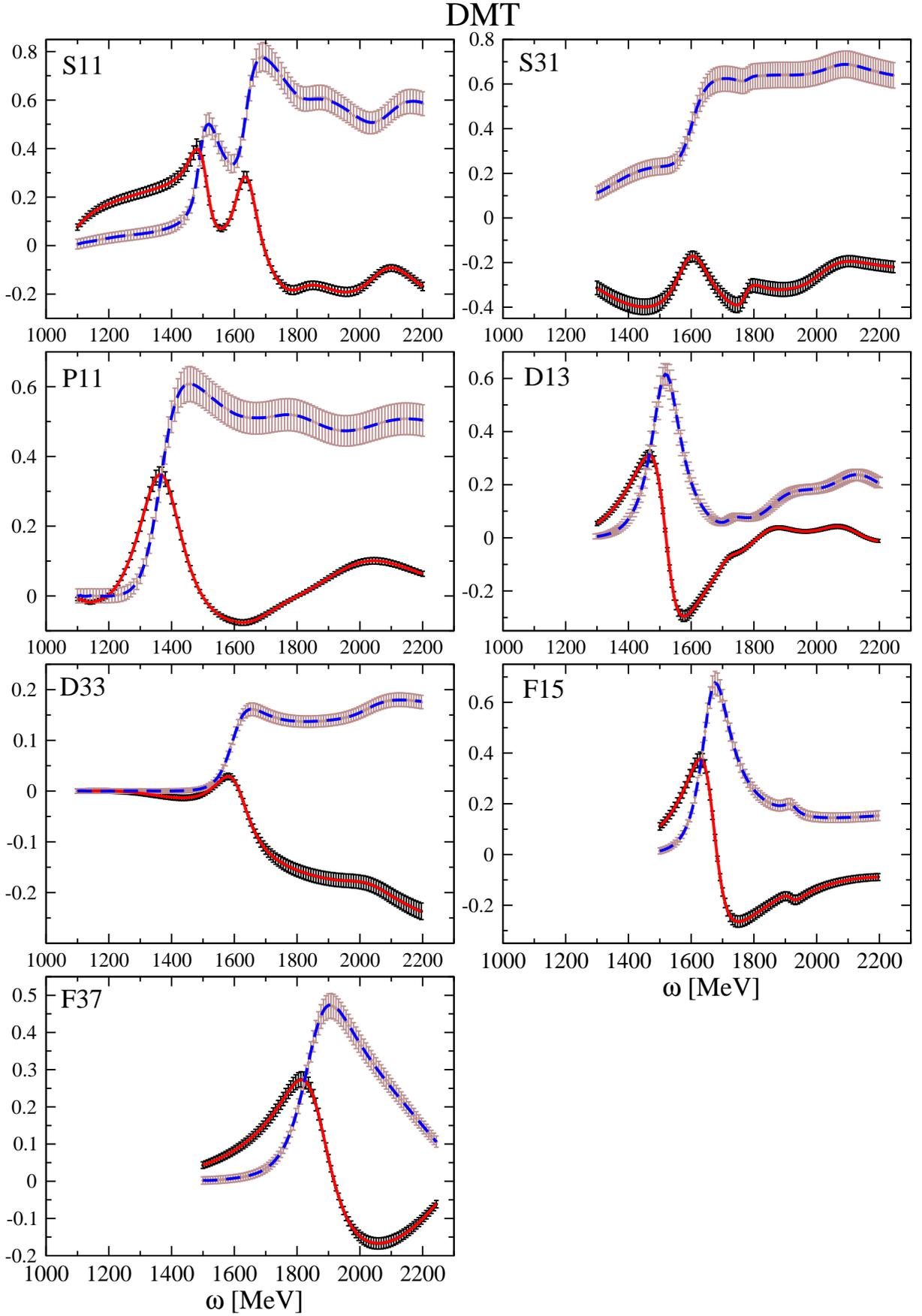}  
\caption{\small(Color online) L+P to partial waves from DMT solution.
 Dashed blue line and solid red line show fit to real and imaginary parts of partial waves, respectively.}
\label{DMT_fig} 
\end{figure*}

Comparison of results from  L+P fit and values from original publications is given in Tables II and III.  Instead of real and 
imaginary parts of residua, in Table II we give their modula $|a_{i}|$ and corresponding phases $\theta_{i}$. $x_{P}, x_{Q}$ 
and $x_{R}$ denote threshold positions {in corresponding Pietarinen series}. As it may be seen from the Tables II and III  
results obtained from L+P agree very well with published values \cite{GWU,GWU1}, and \cite{DMT0,DMT,DMT1}. The quality of the 
fit to the input is given in Figs.~(\ref{Fa02_fig})-(\ref{DMT_fig}).

 \begin{table*}[h!]
 \caption{Comparison of GWU/SAID ED pole positions with values obtained by L+P method.  Table is given in MeV units, and 
$\Gamma _i= - 2 \, \omega_i$. \\ }
\label{GWU/SAID ED} %
\begin{tabular}{||c|c|cccc|cccc||c|c|c|c||}
\hline 
PW  & Solution  & $M_{1}$  & $\Gamma_{1}$  & $|a_{1}|$  & $\theta_{1}^{0}$  & $M_{2-4}$  & $\Gamma_{2-4}$  & $|a_{2-4}|$  & 
$\theta_{2-4}^{0}$  & $x_{P}$/10$^3$    & $x_{Q}$/10$^3$  & $x_{R}$/10$^3$   & $10^{2}\chi_{R}^{2}$\tabularnewline
\hline 
\hline 
 & FA02 \cite{GWU}  & \textbf{1526}  & \textbf{130}  & \textbf{33}  & \textbf{14}  & \textbf{1653}  & \textbf{182}  & 
\textbf{69}  & \textbf{-55} &  &  &  &   \multicolumn{1}{c|}{}\tabularnewline
\cline{2-14} 
\multirow{5}{*}{S$_{11}$ } & FA02 L+P  & 1518  & 121  & 17  & -32  & 1656  & 182  & 68  & -39  & -60.1  & 1.077  & 1.471  & 
0.32\tabularnewline
\cline{2-14} 
 & SP06 \cite{GWU}  & \textbf{1502}  & \textbf{95}  & \textbf{16}  & \textbf{-16}  & \textbf{1648}  & \textbf{80}  & 
\textbf{14}  & \textbf{-69} &  &  &   & \tabularnewline
\cline{2-14} 
 & SP06 L+P  & 1509  & 96  & 15  & -21  & 1645  & 80  & 14  & -80  & -29.5 & 1.077 & 1.479  & 
\multirow{1}{*}{2.90}\tabularnewline
\cline{2-14} 
 & \multirow{2}{*}{WI08 \cite{GWU1} } & \textbf{1499}  & \textbf{98}  & \textbf{-}  & \textbf{-}  & \textbf{1647}  & 
\textbf{84}  & \textbf{-}  & \textbf{-} &  &  &   & \tabularnewline
 &  &   &   &   &   & \textbf{1666}  & \textbf{520}  & \textbf{-}  & \textbf{-} & \multirow{1}{*}{} &  &   & \tabularnewline
\cline{2-14} 
 & \multirow{2}{*}{WI08 L+P } &  1504 & 78  & 11  & -60  & 1644  &  86  & 17  & -83 &  &  &   &  \tabularnewline 
 &  &   &   &   &   & 1669  & 517  & 419  & -74 & -0.339 & 1.077 &  1.483 & 3.57 \tabularnewline
\hline 
\hline 
 & FA02 \cite{GWU}  & \textbf{1594}  & \textbf{118}  & \textbf{17}  & -\textbf{104}  &   &   &   &  &  &  &    & 
\tabularnewline
\cline{2-14} 
\multirow{4}{*}{S$_{31}$} & FA02 L+P  & 1596  & 112  & 15  & -101  &  &  &  &  & \multicolumn{1}{c}{-59.3}  & 1.077 & 1.183  & 
0.48\tabularnewline
\cline{2-14} 
 & SP06 \cite{GWU}  & \textbf{1595}  & \textbf{135}  & \textbf{15}  & \textbf{-92}  &   &   &   &  &  &  &  & \tabularnewline
\cline{2-14} 
 & SP06 L+P  & 1596  & 133  & 18  & -105  &  &  &  &  & -16.7  & 1.077  & 1.309 & 0.35\tabularnewline
\cline{2-14} 
 & WI08 \cite{GWU1}  & \textbf{1594}  & \textbf{136}  & \textbf{-}  & \textbf{-}  &   &   &   &  &    &  &  & \tabularnewline
\cline{2-14} 
 & WI08 L+P  & 1598  & 130  & 18  & -104  &  &  &  &  & -92.7 & 1.077  & 1.589 & 0.57\tabularnewline
\hline 
\hline 
 & FA02 \cite{GWU}  & \textbf{1357}  & \textbf{160}  & \textbf{36}  & \textbf{-102}  &   &   &   &  &  &  &  &  \tabularnewline
\cline{2-14} 
\multirow{6}{*}{P$_{11}$} & FA02 L+P  & 1354  & 169  & 38  & -98  &  &  &  &  & -100 & 1.077 & 1.202  & 0.66 \tabularnewline
\cline{2-14} 
 & SP06 \cite{GWU}  & \textbf{1359}  & \textbf{162}  & \textbf{38}  & \textbf{-98}  &   &   &   &  &    &  &  &  
\tabularnewline
\cline{2-14} 
 & SP06 L+P  & 1358  & 183  & 53  & -92  &  &  &  &  & -62.1  & \multicolumn{1}{c}{1.077 } & 1.215  & 0.09 \tabularnewline
\cline{2-14} 
 & WI08 \cite{GWU1}  & \textbf{1358}  & \textbf{160}  & \textbf{-}  & \textbf{-}  &   &   &   &   &  &  &  &  \tabularnewline
\cline{2-14} 
 & WI08 L+P  & 1357  & 177  & 47  & -93  &  &  &  &  & -98.9  & 1.077 & 1.202  & 0.07\tabularnewline
\hline 
\hline
 & FA02 \cite{GWU}  & \textbf{1514}  & \textbf{102}  & \textbf{35}  & \textbf{-6}  &   &   &   &  &   &  &  & \tabularnewline
\cline{2-14} 
\multirow{4}{*}{D$_{13}$} & FA02 L+P  & 1513  & 101  & 34  & -9  &   & & & & -67.4 & 1.077  & 1.222  & 0.85 \tabularnewline
\cline{2-14} 
 & SP06 \cite{GWU}  & \textbf{1515}  & \textbf{113}  & \textbf{38}  & \textbf{-5}  &  &   &   &  &   &  &  &  \tabularnewline
\cline{2-14} 
 & SP06 L+P  & 1515  & 113  & 38  & -6  &  &  &  &  & -50.1 & 1.077  & 1.216& 0.57 \tabularnewline
\cline{2-14} 
 & WI08 \cite{GWU1}  & \textbf{1515}  & \textbf{110}  & \textbf{-}  & \textbf{-}  &   &   &    &  &  &  &  & \tabularnewline
\cline{2-14} 
 & WI08 L+P  & 1515  & 111  & 38  & -5  &  &  &  &  & -81.1 & 1.077 & 1.169  & 0.15 \tabularnewline
\hline 
\hline 
 & FA02 \cite{GWU}  & \textbf{1617}  & \textbf{226}  & \textbf{16}  & \textbf{-47}  &   &   &   &  &  &  &  &  \tabularnewline
\cline{2-14} 
\multirow{4}{*}{D$_{33}$} & FA02 L+P  & 1618  & 227  & 16  & -47  &  &  &  &  & -27.3 & 1.077  & 1.204 & 0.008 \tabularnewline
\cline{2-14} 
 & SP06 \cite{GWU}  & \textbf{1632}  & \textbf{253}  & \textbf{18}  & \textbf{-48}  &   &   &   &    &  &  &  &  
\tabularnewline
\cline{2-14} 
 & SP06 L+P  & 1635  & 251  & 18  & -37  &  &  &  &  & -54.1 & 1.077  & 1.198 & 0.009 \tabularnewline
\cline{2-14} 
 & WI08 \cite{GWU1}  & \multicolumn{4}{c|}{no results    } &   &  &   &  &  &  &    & \tabularnewline
\cline{2-14} 
 & WI08 L+P  &  &  &  &  &  &  &  &  &  &   &  & \tabularnewline
\hline 
\hline 
 & FA02 \cite{GWU}  & \textbf{1678}  & \textbf{120}  & \textbf{43}  & \textbf{1}  & \textbf{1779}  & \textbf{248}  & 
\textbf{47}  & \textbf{-61} &  &   &  & \tabularnewline
\cline{2-14} 
\multirow{4}{*}{F$_{15}$} & FA02 L+P  & 1679  & 118  & 42  & -5  & 1779  & 245  & 31  & -84  &  1.032 & 1.077  & 1.549  & 0.64 
\tabularnewline
\cline{2-14} 
 & SP06 \cite{GWU}  & \textbf{1674}  & \textbf{115}  & \textbf{42}  & \textbf{-4}  & \textbf{1785}\footnote{While writing this 
paper we have detected a typo in original GWU publication for the second F$_{15}$ resonance. So, in the present publication, 
the original \emph{erroneous} values of 1807 + $i$ 109 have been replaced with corrected values already given on the web-page 
\cite{GWU}. Erratum will follow shortly.}   & \textbf{244}\footnote{See Footnote a.}  & \textbf{60}  & \textbf{-67} &  &  &   & 
\tabularnewline
\cline{2-14} 
 & SP06 L+P  & 1673  & 116  & 43  & -14  & 1776  & 226  & 24  & -98  & -8.99 & 1.077 & 1.301& 0.08 \tabularnewline
\cline{2-14} 
 & WI08 \cite{GWU1}  & \textbf{1674}  & \textbf{114}  & \textbf{-}  & \textbf{-}  & \textbf{1779}  & \textbf{276}  & \textbf{-}  
& \textbf{-} &  \tabularnewline
\cline{2-14} 
 & WI08 L+P  & 1675  & 115  & 44  & -8  & 1776  & 233  & 34  & -99  & -51.7  & 1.077 & 1.726 & 0.28\tabularnewline
\hline 
\hline 
 & FA02 \cite{GWU}  & \textbf{1874}  & \textbf{236}  & \textbf{57}  & \textbf{-34} &   &   &    &  &  &  &  & \tabularnewline
\cline{2-14} 
\multirow{4}{*}{F$_{37}$} & FA02 L+P  & 1874  & 236  & 55  & -35  &  &  &  &  & -14.9 & 1.077  & 1.739 & 0.04 \tabularnewline
\cline{2-14} 
 & SP06 \cite{GWU}  & \textbf{1876}  & \textbf{227}  & \textbf{53}  & \textbf{-31}  &   &   &   &  &    &  &  & \tabularnewline
\cline{2-14} 
 & SP06 L+P  & 1876  & 226  & 53  & -31  &  &  &  &  & -32.8 & 1.077  & 1.137  & 0.06\tabularnewline
\cline{2-14} 
 & WI08 \cite{GWU1}  & \textbf{1883}  & \textbf{230}  & \textbf{-}  & \textbf{-}  &   &   &    &  &  &  &  & \tabularnewline
\cline{2-14} 
 & WI08 L+P  & 1874  & 227  & 55  & -35  &  &  &  &  & -37.8  & 1.077 & 1.736 & 0.03\tabularnewline
\hline 
\end{tabular}
\end{table*}

\begin{table*}[h!]
 \caption{Comparison of DMT ED pole positions with values obtained by L+P method.  Table is given in MeV units, and $\Gamma _i= 
- 2 \, \omega_i$. \\ }
\label{DMT ED} 
\begin{tabular}{||c|c|cccc|cccc||c|c|c|c||}
\hline 
PW  & Solution  & $M_{1}$  & $\Gamma_{1}$  & $|a_{1}|$  & $\theta_{1}^{0}$  & $M_{2-4}$  & $\Gamma_{2-4}$  & $|a_{2-4}|$  & 
$\theta_{2-4}^{0}$  & $x_{P}$/10$^3$    & $x_{Q}$ /10$^3$ & $x_{R}$/10$^3$   & $10^{2}\chi_{R}^{2}$\tabularnewline
\hline 
\hline 
 \multirow{6}{*}{S$_{11}$}& DMT \cite{DMT0,DMT}  & \textbf{1499}  & \textbf{78}  & \textbf{14}  & \textbf{-45}  & \textbf{1631}  
& \textbf{120}  & \textbf{35}  & \textbf{-83} & \multicolumn{4}{c||}{} \tabularnewline
&   & \textbf{}  & \textbf{}  & \textbf{}  & \textbf{}  & \textbf{1733}  & \textbf{180}  & \textbf{16}  & \textbf{-29} & 
\multicolumn{4}{c||}{} \tabularnewline
& DMT \cite{DMT1}  & \textbf{}  & \textbf{}  & \textbf{}  & \textbf{}  & \textbf{2027}  & \textbf{180}  & \textbf{23}  & 
\textbf{-150} & \multicolumn{4}{c||}{} \tabularnewline
\cline{2-14} 
 & DMT L+P  &  1500 &  76 &  13.4 &  -46 & 1636  & 99  & 22  & -94  &      \multicolumn{4}{c||}{}  \tabularnewline
 &  &       &     &       &      & 1810  & 164 & 9.6 & -176 &      \multicolumn{4}{c||}{} \tabularnewline
 \cline{11-14}
 &  &       &     &       &      & 2077  & 220 & 22.5& -122  & 1.0 & 1.077 & 1.486  & \multirow{1}{*}{0.6}\tabularnewline
\hline 
\hline 
 \multirow{5}{*}{S$_{31}$}& DMT \cite{DMT0,DMT}  & \textbf{1598}  & \textbf{148}  & \textbf{23}  & \textbf{-98}  & 
\textbf{1774}  & \textbf{72}  & \textbf{3.8}  & \textbf{-181} & \multicolumn{4}{c||}{} \tabularnewline
&  & \textbf{}  & \textbf{}  & \textbf{}  & \textbf{}  & \textbf{1984}  & \textbf{254}  & \textbf{26}  & \textbf{-170} & 
\multicolumn{4}{c||}{} \tabularnewline
\cline{2-14} 
 & DMT L+P  &  1597 &  140 &  21 &  -104 & 1771  & 69  & 2.2  & -172  &      \multicolumn{4}{c||}{}  \tabularnewline
 \cline{11-14}
 &  &       &     &       &      & 2040  & 195 & 7 & -109   & -11.476 & 1.077 & 1.739 &  \multirow{1}{*}{0.2}  \tabularnewline
 \hline 
 \multirow{4}{*}{P$_{11}$}& DMT \cite{DMT0,DMT}  & \textbf{1371}  & \textbf{190}  & \textbf{50}  & \textbf{-79}  & 
\textbf{1746}  & \textbf{368}  & \textbf{11}  & \textbf{-54} & \multicolumn{4}{c||}{} \tabularnewline
& DMT \cite{DMT1} & \textbf{}  & \textbf{}  & \textbf{}  & \textbf{}  & \textbf{1997}  & \textbf{458}  & \textbf{56}  & 
\textbf{-145} & \multicolumn{4}{c||}{} \tabularnewline
\cline{2-14} 
 & DMT L+P  &  1370 &  190 &  50 &  -81& 1763  & 235  & 5  & -56  &     \multicolumn{4}{c||}{}  \tabularnewline
 \cline{11-14}
 &  &       &     &       &      & 2015  & 467 & 36 & -99 & 0.699 & 1.077 & 1.537  & \multirow{1}{*}{0.05} \tabularnewline
 \hline 
\hline 
 \multirow{4}{*}{D$_{13}$}& DMT \cite{DMT0,DMT}  & \textbf{1515}  & \textbf{120}  & \textbf{40}  & \textbf{-7}  & \textbf{1718}  
& \textbf{96}  & \textbf{2.8}  & \textbf{-91} & \multicolumn{4}{c||}{} \tabularnewline
&   & \textbf{}  & \textbf{}  & \textbf{}  & \textbf{}  & \textbf{1854}  & \textbf{214}  & \textbf{16}  & \textbf{-96} & 
\multicolumn{4}{c||}{} \tabularnewline
& DMT \cite{DMT1}  & \textbf{}  & \textbf{}  & \textbf{}  & \textbf{}  & \textbf{2099}  & \textbf{216}  & \textbf{13}  & 
\textbf{-58} & \multicolumn{4}{c||}{} \tabularnewline
\cline{2-14} 
 & DMT L+P  &  1517 &  120 &  40 &  -5& 1721  & 89  & 2.1  & -76  &      \multicolumn{4}{c||}{}  \tabularnewline
 &  &       &     &       &      & 1858& 228 & 15& -87&      \multicolumn{4}{c||}{} \tabularnewline
 \cline{11-14}
 &  &       &     &       &      & 2101 & 231& 14& -49  & 1.00 & 1.077 & 1.266  & \multirow{1}{*}{0.32}\tabularnewline
\hline 
\hline 

 \multirow{2}{*}{D$_{33}$}& DMT \cite{DMT0,DMT,DMT1}  & \textbf{1604}  & \textbf{142}  & \textbf{9.4}  & \textbf{-63}  & 
\textbf{2042}  & \textbf{254}  & \textbf{4.84}  & \textbf{-75} & \multicolumn{4}{c||}{} \tabularnewline
\cline{2-14} 
 & DMT L+P  &  1605 &  141 &  9.3 &  -63& 2023  & 241  & 4  & -93 &  0.623 & 1.077 & 1.324 & \multirow{1}{*}{0.06} 
\tabularnewline
 \hline 
\hline 
 \multirow{2}{*}{F$_{15}$}& DMT \cite{DMT0,DMT}  & \textbf{1664}  & \textbf{114}  & \textbf{38}  & \textbf{-26}  & 
\textbf{1919}  & \textbf{52}  & \textbf{1.0}  & \textbf{15} & \multicolumn{4}{c||}{} \tabularnewline
\cline{2-14} 
 & DMT L+P  &  1664 & 114 &  38 &  -26& 1920  & 52  & 1.0  & 16 &   0.7 & 1.077 & 1.225 & \multirow{1}{*}{0.02}  
\tabularnewline
\hline  
\hline 
\multirow{2}{*}{F$_{37}$} &  DMT \cite{DMT0,DMT}    & \textbf{1858}  & \textbf{208}  & \textbf{43}  & \textbf{-48}  &   &   &   
&  &    \multicolumn{4}{c||}{} \tabularnewline
\cline{2-14} 
 & DMT  L+P  & 1858  & 207  & 43  & -49 &  &  &  &  & -3.999 & 1.077  & 1.223  & 0.48\tabularnewline
\hline 
\hline 
\end{tabular}
\end{table*}

\clearpage
\section{Results and Discussion}

\subsection{General considerations}

By using Laurent expansion near real axes, and Pietarinen series to constrain unknown regular part, we have been able to 
extract pole parameters \emph{in all test cases} where the pole parameters have been known by using other methods without 
exemption, and our value \emph{corresponded to the know input fairly well}.
\subsubsection{Toy-model}
First we tested the toy-model where the analytic structure was known, and relatively simple. The agreement given in Table 
\ref{tb1:Toy model parameters} is below 1~\%. 
\subsubsection{GWU/SAID ED solutions}
To illustrate the applicability of the method to situations where the analytic structure of the input is much more complicated, 
and we are approximating the complicated input analytic function with simpler one, we have tested two sets of ED amplitudes; 
GWU/SAID \cite{GWU,GWU1}, and DMT \cite{DMT}. We have compared only those partial waves for which the input pole parameters are 
explicitly published. 

Results for GWU/SAID amplitude are give in Table~\ref{GWU/SAID ED}, and in Figs.~(\ref{Fa02_fig})-(\ref{Wi08_fig}).  The 
GWU/SAID fit uses a Chew-Mandelstam K-matrix model and the resulting T-matrix poles are not generally included as explicit 
K-matrix
poles. In addition, the model employs complex branch points for the opening $\pi \Delta$ and $\rho N$ channels.  
Given these complications, we anticipated that the analytic structure of the regular part could be rather complicated, and we 
would have some problems in fitting these amplitudes within our method. This expectation is 
confirmed. In spite of an almost ideal ability of our simpler background to reproduce the more complicated input, demonstrated 
by extremely good fits given in  Figs.~(\ref{Fa02_fig})-(\ref{Wi08_fig}), and low $\chi^2$ parameter given in 
Table~\ref{GWU/SAID ED}, some differences in pole parameters should and do occur. However, we  state:
\begin{itemize}
\item the quality of the fit is ideal (see Figs.~\ref{Fa02_fig}, \ref{Sp06_fig} and \ref{Wi08_fig})
\item  we find an identical number of poles as the GWU/SAID group
\item  we are able to follow the significant change of pole positions, due to slight changes of input, as reported by the 
GWU/SAID group (the input for the S11 partial wave is only modestly modified at higher energies between the FA02 and SP06 
solutions, but the change in width of the S11(1650) resonance is noticeable: from 182 MeV for FA02 to 80 for SP06. We reproduce 
this finding.)
\item our agreement for the lowest partial waves is in principle better than 10~\% for lower resonances, and only slightly 
worse for higher ones
\item our L+P method disagreed with the GWU/SAID pole position published for the second F15 resonance, and this revealed a 
missprint in the original publication (see footnote in Table~\ref{GWU/SAID ED}).
\end{itemize}

\subsubsection{DMT ED solutions}
As the DMT model is a T-matrix model, we expect that its analytic
structure will be much easier to reconstruct for the L+P model, so
we expect better results. It is based on a $\pi N$ meson exchange
model and describes $\pi N$ phase shifts and inelasticity parameters
in all the partial waves up to the F waves and energies of
$W=2$~GeV. It is a dynamical model with $\pi N$, $2\pi N$ and $\eta
N$ coupled channels, leading to a simpler branch point structure than
the GWU/SAID K-matrix approach or more sophisticated dynamical
models with many more coupled channels and complex branch
points~\cite{Ronchen2012}. Most resonances are included as bare
resonances which get dressed by the dynamically calculated
self-energies, and the pole positions and residues can be calculated
by analytical continuation into the complex region. In addition,
like in all dynamical models, dynamically generated poles can be
found, most of them far away form the physical axis. The results and
the quality of the fit is shown in Table~\ref{DMT ED}. We state:
\begin{itemize}
\item the quality of the fit is ideal (see Fig.~\ref{DMT_fig})
\item we find the same number of poles as DMT group~\cite{DMT}
\item our L+P method disagrees with the DMT published pole position for the second P11 resonance,
where we observe a $30\%$ deviation in the pole width and a $50\%$
deviation in the residue. However, this pole has the smallest
''normalized residue'', defined as $R_{norm}=Res/(\Gamma/2)$, from
all resonances in Table~\ref{DMT ED}. Its value is only $6\%$
whereas the largest one in the table is found for the $D_{13}(1520)$
with $67\%$. Also small values are found for the third $S_{11}$
($18\%$), the second $S_{31}$ ($11\%$) and second, third and fourth
$D_{13}$ ($6\%$, $15\%$ and $12\%$, respectively). Quantitatively,
we observe larger deviations as smaller the reduced residues get.
This quantity is a very good measure for the strength and importance
of a resonance, the maximum value among all nucleon resonances is
obtained for the $\Delta(1232)$ with $100\%$.
\end{itemize}

\subsection{GWU/SAID P$_{11}$ poles in ED and single energy solutions (SES)}

\begin{figure*}[!t]
\includegraphics[width=14cm]{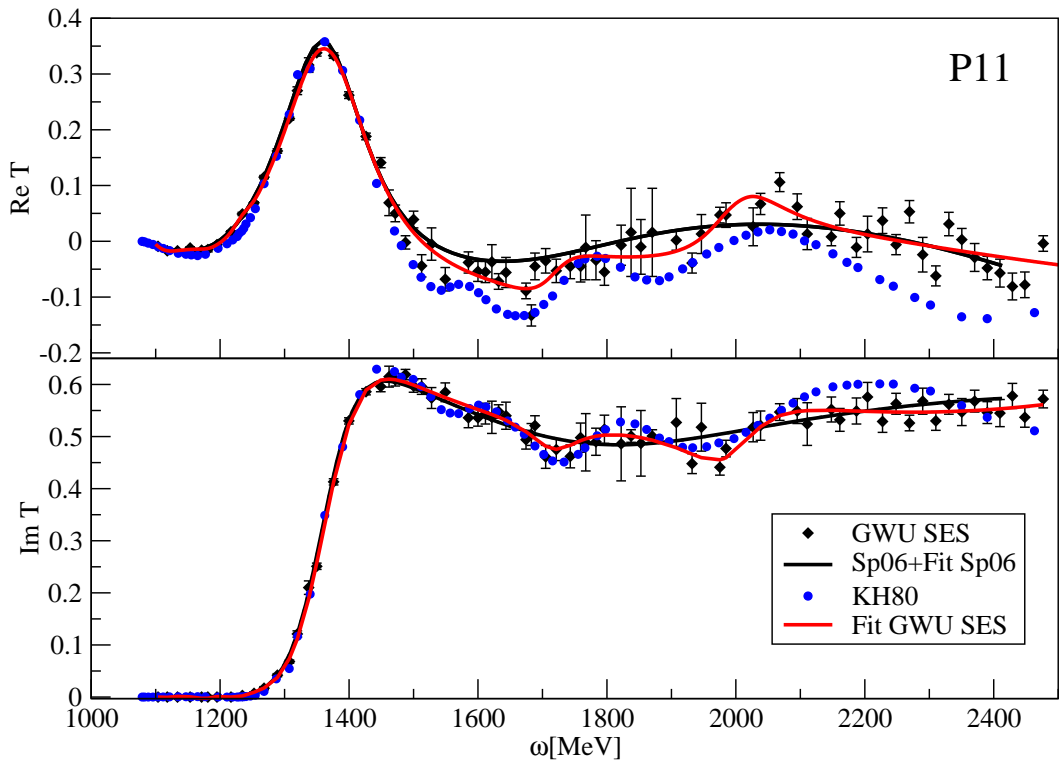}  
\caption{\small{(Color online) Comparison beetween GWU-SES, GWU-SP06 and Karlsruhe-Helsinki KH80 \cite{Hoehler84} solution  for 
P11 partial wave.}}
\label{Wi08_slika} 
\end{figure*}

In order to demonstrate the importance and novelty of extracting poles from amplitudes more closely associated with 
experimental data using L+P method, we have extracted poles from the $P_{11}$ partial wave of the GWU/SAID SES, and have 
compared these with 
results using the ED GWU/SAID fit SP06. The results are interesting. However, before discussing this fit, it is
useful to recall how the SES fits differ (and are related to) the ED fits. 

The SES amplitudes come from fits
to data within narrow windows of energy. The energy variation over each range is determined by the underlying
ED fit. However, as each fit is done independently, there is no smoothness constraint applied to the set of 
resulting SES. Dispersion relation constraints are also omitted in these fits, which were done initially to
search for systematic variations, possibly signaling missing structure. Having obtained these SES, there is
an element of subjectivity in assessing whether the variations are random or indicative of missing (possibly
resonant) structure. In the case of the P$_{11}$ amplitude, other analyses (both multi-channel and elastic)
have suggested the existence of several P$_{11}$ states, whereas the GWU/SAID fits find only the single (Roper)
resonance $N(1440)$. As no explicit pole is added in the fit, the Roper has been found in a search of the
complex energy plane. 

The P$_{11}$ amplitude is complicated by the existence of the $\pi \Delta$ branch point very near the first
resonance pole. The amplitude, plotted in an Argand diagram, also stays near its center for energies spanning
the whole resonance region beyond the Roper, which hinders the determination of phase (looping) behavior. 
Several attempts have been made to add structure~\cite{GWU1,VPI} either by inserting resonances by hand,
adding soft pseudo-data constaints based on the KH80~\cite{Hoehler84} and 
CMB~\cite{CMB} amplitudes, or adding an extra explicit pole
to each partial wave via the Chew-Mandelstam K-matrix. In the latter attempt, a second pole was extracted but
its position was not comparable to other determinations. A comparison of the GWU/SAID SES and the much older
KH80 fit is suggestive as KH80 finds a second resonance and seems to follow all the structure seen in the
SES.

We can fit both the ED and SES solutions. 
The quality of the fit is shown in Fig.~(\ref{Wi08_slika}). However, we need three poles for GWU-SES, but need \emph{only one} 
pole for SP06.  We have tried fitting SP06 with two poles, but the the second pole turned out to be completely undetermined. 
However, when fitting GWU-SES as input, the existence of the second pole is definitely established, and the existence of the 
third pole is very likely.  These results are very similar to what GWU group claims:  P$_{11}(1710)$ is not found in the 
complex plane of the ED analysis. 
\\ \noindent
Our results are: 
\begin{itemize}
\item  GWU-SP06  \\
\hspace*{1.8cm}$P_{1}^{\rm SP06}=1.358-i\:0.0915$
\item GWU-SES 
\begin{eqnarray*}
P_{1} & = & 1.362-i\:0.0895,\\
P_{2} & = & 1.716-i\:0.0495,\\
P_{3} & = & 1.999-i\:0.0715.
\end{eqnarray*}
\end{itemize} 
Using the L+P method allows us to see P$_{11}(1710)$ and P$_{11}(2100)$ in the GWU-SES, while GWU/SAID energy dependent 
analysis 
does not contain this additional structure. It would be interesting to see if, by adding further explicit poles, this result
could be found also in the GWU/SAID ED approach.

\begin{acknowledgements}
The authors of this paper are especially grateful to Jambul Gegelia, University of Mainz, for extremely valuable comments 
regarding the radius of convergence of the L+P model what helped to clarify a lot of misunderstandings which occurred during 
and after the first presentation of the model at the Workshop:"PWA Tools in Hadron Spectroscopy", in Mainz in February this 
year.
\end{acknowledgements}

\end{document}